\documentclass[a4paper]{jpconf}
\usepackage{graphicx}
\usepackage{iopams}
\usepackage{amsmath}
\usepackage{cite}

\newcommand{\gbar}{\bar{g}}

\renewcommand{\d}{\mathrm{d}}
\renewcommand{\L}{\mathcal{L}}

\begin{document}
\title{Potentially healthy time evolution in interacting $p$-form fields}

\author{Fethi M. Ramazano\u{g}lu\orcid{0000-0003-3075-1457}}
\address{Department of Physics, Ko\c{c} University, Rumelifeneri Yolu, 34450 Sar{\i}yer, \.{I}stanbul, T\"{u}rkiye}
\ead{framazanoglu@ku.edu.tr}

\author{K{\i}van\c{c} \.I. \"Unl\"ut\"urk\orcid{0000-0002-0289-9764}}
\address{Department of Electrical and Electronics Engineering, Turkish-German University, 34820 Beykoz, \.{I}stanbul, T\"{u}rkiye}
\ead{kivanc.unluturk@tau.edu.tr}

\begin{abstract}
It was recently discovered that many self-interacting vector fields can have healthy time evolution when field amplitudes are small, however, further dynamics becomes impossible if the norm of the field reaches a certain finite value. This result was also generalized to higher form fields using the fact that vector fields can also be viewed as 1-form fields. However, other recent work demonstrated that some exceptional vector field theories do not suffer from such problems if the self-interaction term is chosen in a specific way. Here, we study whether a similar exception exists for $p$-form fields in general. We show that, in 4 spacetime dimensions, 3-form fields with self interaction can indeed have healthy time evolution similarly to vectors. However, the analogous coupling terms still lead to loss of hyperbolicity for 2-form fields. We study the reasons for these differences, and comment on directions to generalize our findings. We also demonstrate that eliminating the mass term, one of the proposals to build healthy self-interacting vector field theories, moves the singularity to other parts of the equations of motion, hence, most likely does not provide well-posed time evolution.
\end{abstract}

\emph{Dedicated to Metin Gürses on the occasion of his becoming an emeritus professor after 60 years of research
and teaching.}
\section{Introduction}
Time evolution is an essential ingredient of our understanding of nature, hence we desire our theories to reflect it in an appropriate manner. Even though it is taken for granted in many theory building exercises, recent work has shown that some of the simplest vector field theories lack an indefinite well-posed time evolution. Namely, if a vector field theory has generic self-interaction terms, time evolution becomes impossible after a finite duration in a phenomenon called \emph{loss of hyperbolicity}~\cite{EspositoFarese2010, Mou2022, Clough2022, Coates2022, Barausse2022, Coates2023a, Coates2023,Unluturk2023,Doneva2024}. Moreover, $p$-form fields, that is, totally antisymmetric tensor fields, are also known to suffer from similar problems due to their mathematical similarities to vectors, which have a one-to-one correspondence with 1-form fields~\cite{Unluturk2024}. This has important consequences since such fields play important roles in many theories of gravity, cosmology and high energy physics.

Even though the aforementioned problems occur for generic coupling terms, some recent studies showed that there are ``exceptional'' cases where time evolution can continue forever for vector fields~\cite{Rubio2024,Banerjee:2025fph}. This is an interesting point, since it might provide guidance in theory building. Studying the underlying reasons for better behavior can provide tools for building more physically-relevant models of the universe. Our main task in this paper is exploring such ideas for higher ($p>1$) $p$-form fields, and investigating whether similar coupling terms can provide well-posed theories in these cases as well.

Our main finding is that results on well-posed self-interacting vector field theories can be useful for $p$-form fields as well, but in a limited manner. Some $p$-form fields can have pathology-free time evolution when we introduce modified self-interaction terms which are inspired from the vector field studies, whereas others are still problematic in all cases we have studied. This follows our previous understanding of loss of hyperbolicity in higher form field theories. When there is a pathology, it is known that some theories feature it very similarly to vector fields, whereas the time evolution singularity is quite subtler in others. We will specifically present these points in the examples of 2- and 3-form fields in 4 spacetime dimensions: the latter behaves in close analogy to vectors and can have healthy time evolution for certain couplings, whereas the former does not.

A second finding of ours is about the effectiveness of the proposals to obtain well-posed time evolution in self-interacting theories. We consider two of these: changing the self-interaction potential and removing the mass term. Further studies are needed for a complete understanding, but we show that the latter proposal does not lead to the purported result. The massless limit of self-interacting $p$-form fields still suffer from time evolution problems. This approach simply moves the singularity from one term to another as far as our methods indicate.

In Sec.~\ref{sec: self-interacting p-form} we will overview the literature for loss of hyperbolicity in vectors and higher form fields. In Sec.~\ref{sec: 1-form loss of hyperbolicity exception}, we explain the exceptional theories where time evolution is purported to be healthy,  and discuss the problems with the idea of removing the mass term. In  Sec.~\ref{sec: higher form loss of hyperbolicity exception}, we discuss how the ideas from the healthy exceptional vector field theories translate to higher form fields. Sec.~\ref{sec: conclusion} contains our conclusions and outlook.

We work with Cartesian coordinates in flat Minkowski spacetime with the mostly plus metric signature and $c=1$, so $g_{\mu\nu}=\eta_{\mu\nu}$. We still keep the general metric notation, since most of our results can be generalized to curved spacetime, but we adopted the flat spacetime restriction to limit computational clutter in this initial study. Antisymmetrization on tensor indices is denoted by square brackets ``[ ]", and indices to be excluded from antisymmetrization are placed between vertical lines. For instance: ``$[\mu|\alpha\beta|\nu\gamma]$" represents antisymmetrization on only $\mu$, $\nu$ and $\gamma$. We keep our formulas general for any spacetime dimension $D$ at first, but many of our later results are specific to $D=4$, which will be explicitly stated. Lastly, the Hodge dual of a $p$-form $B_{\mu_1 \cdots \mu_p}$ is the $(D{-}p)$-form
\begin{equation}
\tilde{B}_{\nu_1 \cdots \nu_{D-p}} = \frac{1}{p!} \epsilon_{\mu_1 \cdots \mu_p \nu_1 \cdots \nu_{D-p}} B^{\mu_1 \cdots \mu_p}.
\end{equation}

\section{Hyperbolicity in self-interacting form field theories}
\label{sec: self-interacting p-form}

\subsection{Loss of hyperbolicity in 1-forms}
\label{sec: 1-form loss of hyperbolicity}

1-forms are the setting where the time evolution problems we are interested in were discovered~\cite{EspositoFarese2010, Coates2022,Clough2022, Mou2022}. 1-forms have a one-to-one correspondence with vector fields, so we will use the terms interchangeably. We mainly follow the vector field discussion of Ref.~\cite{Coates2022} in this subsection.

Consider the self-interacting Lagrangian for the 1-form field $A$
\begin{equation}
\label{eq: 1-form Lagrangian}
\L = -\frac{1}{4}F_{\mu\nu}F^{\mu\nu} - \frac{1}{2}m^2 A_\mu A^\mu - V(A),
\end{equation}
where $F_{\mu\nu} = \partial_\mu A_\nu - \partial_\nu A_\mu = 2\partial_{[\mu} A_{\nu]}$ is the \emph{field strength tensor}, $m$ is the mass parameter and $V$ is the self interaction potential which depends on $A$ and the metric.\footnote{Derivative terms are also possible in general as studied in Ref.~\cite{Unluturk2023}, but we will not be discussing such theories here.}

Arguably, the simplest self-interaction term is the quartic one
\begin{equation}
\label{eq: 1-form quartic potential}
    V = \frac{\lambda m^2}{4}\left(A_\mu A^\mu\right)^2,
\end{equation}
where $\lambda$ is a coupling constant, which leads to the field equation
\begin{equation}
\label{eq: 1-form eom}
\partial_\mu F^{\mu\nu} = m^2 z A^\nu,
\end{equation}
where
\begin{equation}
\label{eq: 1-form z}
z=1 + \lambda A_\mu A^\mu.
\end{equation}

If we use the antisymmetry of $F^{\mu\nu}$ after applying $\partial_\nu$ to Eq.~\eqref{eq: 1-form eom}, we reach 
\begin{equation}
\label{eq:generalized_lorenz_1form}
\partial_\mu(z A^\mu) = 0.
\end{equation}
This is called the \emph{generalized Lorenz condition}. We emphasize that it is not a gauge choice, a self-interacting vector field does not have gauge freedom. Rather, this is a necessary constraint that has to be followed by the field due to its very nature. 

Despite its difference, the generalized Lorenz condition can be useful in a similar manner to the gauge conditions, which help to show that the electromagnetic field equations can be cast in the form of a wave equation. Namely, relatively straightforward manipulations show that the field equation~\eqref{eq: 1-form eom} can be recast as~\cite{Coates2022}
\begin{equation}
\label{eq: 1-form eom wavelike form}
\square A^\nu + \partial^\nu \left(\frac{1}{z} A^\mu \partial_\mu z \right) = m^2 z A^\nu,
\end{equation}
with $\square = \partial^\mu \partial_\mu$. This equation is not manifestly a wave equation,\footnote{Curiously, the equations can be recast exactly as a wave equation in 1+1 dimensions~\cite{Coates2022}, but here we will present the more general case.} and requires a \emph{principal symbol} analysis.

The principal symbol essentially analyzes the behavior of the Fourier modes of the theory after linearizing it on an arbitrary, fixed background solution, where we freeze all the coefficients of the partial differential equation. One cannot ensure that the theory is healthy if this analysis does not discover any pathologies. However, if there is a time evolution problem in the linearized theory, it is known that the fully nonlinear theory also suffers from the same issues~\cite{Strang1966, Sarbach2012, Kovacs2020}.

The procedure starts with a single mode $A_\mu=\chi_\mu e^{i k_\nu x^\nu}$ inserted to the principal part, the terms with the highest derivatives of the field equation. After some algebra, this term reduces to the form $\mathcal{P}(k)^\mu_{\phantom{\mu}\nu} \chi_\mu$, $\mathcal{P}$ being the principal symbol. For Eq.~\eqref{eq: 1-form eom wavelike form},
\begin{equation}
\mathcal{P}(k)^\mu_{\phantom{\mu}\nu} = k^2 \delta^\mu_\nu + \frac{2\lambda}{z} A^\alpha A^\mu k_\alpha k_\nu
\end{equation}
where $A^\rho$ is the background field value and $k^2 = k^\alpha k_\alpha$. 

Loss of hyperbolicity occurs when the frequency $k^0$ becomes imaginary, which in turn leads to exponentially growing modes. Moreover, the growth rate typically diverges as the wavenumber of the mode approaches infinity. This means, an arbitrarily small contribution from such a mode would lead to arbitrarily fast growth, rendering time evolution ill-posed.

The principal symbol reveals the above behavior since it provides the dispersion relation of the modes via $\det(\mathcal{P}(k)^\mu_{\phantom{\mu}\nu}) = 0$. For our case, in $D$ spacetime dimensions
\begin{equation}
\det(\mathcal{P}(k)^\mu_{\phantom{\mu}\nu}) = \left(g_{\mu\nu} k^\mu k^\nu\right)^{D-1} \frac{\left(\gbar_{\alpha\beta} k^\alpha k^\beta\right)}{z} = 0,
\end{equation}
That is, the dispersion of $A_\mu$ depends on two metrics. The spacetime metric $g_{\mu\nu}$, \emph{and} the \emph{effective metric} $\gbar_{\alpha\beta}$,
\begin{equation}
\label{eq: 1-form effective metric}
\gbar_{\alpha\beta} = zg_{\alpha\beta} + 2\lambda A_\alpha A_\beta.
\end{equation}
However, the effective metric has a more complex structure, and its signature is not obvious. If this metric becomes non-Lorentzian, then there are modes with imaginary frequencies, and we can conclude that hyperbolicity is lost for certain values of $A$.

The tool that can inform us about the signature of $\gbar_{\alpha\beta}$ is its determinant $\bar{g}$. The matrix determinant lemma
\begin{equation}
\label{eq: sylvester's det theorem}
\det(M+uv^{\text{T}}) = \left(1 + v^{\text{T}}M^{-1}u\right) \det(M)
\end{equation}
ultimately leads to the key result
\begin{equation}
\label{eq: 1-form det gbar}
\gbar = - \left(1 + \lambda A_\mu A^\mu \right)^{D-1} \left(1 + 3\lambda A_\mu A^\mu \right) = -z^{D-1}(3z-2).
\end{equation}
The sign of $\gbar$ is the same (negative) as that of $g$ when $|\lambda A_\mu A^\mu| \ll 1$, and both metrics have a healthy dispersion relation. However, irrespective of the value or sign of the coupling constant $\lambda$, there are always finite field configurations $A_\mu$ for which $\gbar$ becomes positive. In that case, $\gbar_{\mu\nu}$ becomes non-Lorentzian, imaginary frequencies appear, and time evolution becomes ill-posed. This is loss of hyperbolicity.

When the initial data is healthy and we start from small values of $|A_\mu A^\mu|$, $\left(1 + 3\lambda A_\mu A^\mu \right)$ in Eq.~\eqref{eq: 1-form det gbar} is always the term to cause the problem for the first time, so hyperbolicity is lost when $\lambda A_\mu A^\mu =-1/3$. The picture so far only provides the field values for which time evolution is possible or not. Another important question is whether we ever reach the problematic field values if the initial data is healthy. After all, it is conceivable that the dynamics works in such a way that the pathological configurations are always avoided. This scenario is not what happens, and numerical computations have demonstrated that healthy initial data can evolve into loss of hyperbolicity for any theory parameters $(m,\lambda)$~\cite{Coates2022, Coates2023a, Coates2023}.

\subsection{Loss of hyperbolicity in  higher forms}
\label{sec:higher_forms_general}
We will be closely following Ref.~\cite{Unluturk2024} in this section. A $p$-form field is a completely antisymmetric rank $(0,p)$ tensor field $B_{\mu_1\cdots\mu_p}=B_{[\mu_1\cdots\mu_p]}$. A massive $p$-form field with self interaction terms has the canonical Lagrangian
\begin{equation}
\label{eq: Lagrangian of p-form}
\L = -\frac{1}{2} \frac{1}{\left(p+1\right)!} M^2 -\frac{1}{2} \frac{1}{p!}m^2 B^2 - V(B),
\end{equation}
in analogy to Eq.~\eqref{eq: 1-form Lagrangian}. Here, 
\begin{equation}
    M^2 = M_{\mu_1\cdots\mu_{p+1}}M^{\mu_1\cdots\mu_{p+1}}, \quad
    B^2 = B_{\mu_1\cdots\mu_p}B^{\mu_1\cdots\mu_p},
\end{equation}
$M=\d B$ being the field strength tensor or the \emph{exterior derivative} of $B$
\begin{equation}
    M_{\mu_1\cdots\mu_{p+1}} = \left(p+1\right) \partial_{[\mu_1} B_{\mu_2\cdots\mu_{p+1}]}.
\end{equation}
The equation of motion that follows from Eq.~\eqref{eq: Lagrangian of p-form} is
\begin{equation}
\label{eq: p-form eom for general potential}
\partial_\alpha M^{\alpha\mu_1\cdots\mu_p} = m^2 B^{\mu_1\cdots\mu_p} + p!\frac{\partial V}{\partial B_{\mu_1\cdots\mu_p}}.
\end{equation}
$\partial V/\partial B_{\mu_1\cdots\mu_p}$ is the antisymmetric tensor such that $\delta V = (\partial V/\partial B_{\mu_1\cdots\mu_p}) \delta B_{\mu_1\cdots\mu_p}$ under a variation $\delta B_{\mu_1\cdots\mu_p}$.
Arguably the simplest self interaction potential is
\begin{equation}
\label{eq: p-form simple potential}
    V = \frac{\lambda m^2}{4}\left(\frac{1}{p!}B^2\right)^2.
\end{equation}

The equation of motion for this theory is
\begin{equation}
\label{eq: p-form eom for simplest quartic potential}
\partial_\alpha M^{\alpha\mu_1\cdots\mu_p} = m^2 z B^{\mu_1\cdots\mu_p},
\end{equation}
where 
\begin{equation}
\label{eq: p-form z}
z = 1 + \frac{\lambda}{p!} B^2.
\end{equation}
Note that everything is a close analog of the 1-form, or equivalently the vector, theory so far, which is also reflected in the subsequent discussion.

We can obtain a further generalization of the generalized Lorenz condition by applying $\partial_{\mu_1}$ to Eq.~\eqref{eq: p-form eom for simplest quartic potential} and using the antisymmetry of $M$, obtaining
\begin{equation}
\label{eq: p-form Lorenz condition}
\partial_\alpha \left(zB^{\alpha\mu_2\cdots\mu_p}\right)=0.
\end{equation}
We can then recast the field equation as 
\begin{equation}
\label{eq: p-form eom wavelike form}
    \square B^{\mu_1\cdots\mu_p} + \frac{p}{z} \partial_\alpha \partial^{\left[\mu_1\right|} z B^{\alpha\left|\mu_2\cdots\mu_p\right]}
    + \frac{p}{z} \partial_\alpha z \partial^{\left[\mu_1\right|} B^{\alpha\left|\mu_2\cdots\mu_p\right]} = m^2 z B^{\mu_1\cdots\mu_p}.
\end{equation}
After some lengthy algebra, the dispersion relation is revealed to depend on an effective metric as before, which is given by
\begin{equation}
\label{eq: p-form effective metric}
\gbar_{\alpha\beta} = zg_{\alpha\beta} + \frac{2\lambda}{\left(p-1\right)!} B^{\phantom{\alpha}\mu_2\cdots\mu_p}_{\alpha} B_{\beta\mu_2\cdots\mu_p}.
\end{equation}
This finally brings us to the point where we can analyze the hyperbolicity of these theories. 

In 4 spacetime dimensions, the only nontrivial cases are $p=2$ and $p=3$,\footnote{The 4-form is non-dynamical since $M = \d B$ vanishes in 4 dimensions. If we only have the kinetic term, the gauge invariant 3-form field theory is also non-dynamical in 4 dimensions. However, this is not the case with self interaction, see the appendices of Ref.~\cite{Unluturk2024} for details of the self-interacting 3-form fields.} for which the determinants of the effective metrics $\bar{g}^{(p)}$ are
\begin{align}
\label{eq: 2-form det gbar}
\bar{g}^{(2)} &= - \left[z\left(1 + \frac{3\lambda}{2}B_{\mu\nu}B^{\mu\nu}\right) - \left(\frac{\lambda}{2} B_{\mu\nu}\tilde{B}^{\mu\nu}\right)^2 \right]^2\\
\bar{g}^{(3)} &= - \left(1 + \frac{\lambda}{6} B^2\right) \left(1 + \frac{\lambda}{2} B^2\right)^3.
\end{align}

Let us first discuss $\bar{g}^{(3)}$, which is a relatively more straightforward analog of $\bar{g}$ for 1-form fields in Eq.~\eqref{eq: 1-form det gbar}. Namely, $\bar{g}^{(3)}$ can change its sign when $\lambda B^2 < -2$, which is the signal for loss of hyperbolicity. This case has not been examined in full computational detail so far, but the evidence from $p$-form fields in general strongly indicates that initially healthy 3-form field configurations can dynamically reach a point where loss of hyperbolicity occurs.

The case of $\bar{g}^{(2)}$ is less transparent. This formula has the radical difference from the others so far that, it is manifestly non-positive. Hence, the determinant of the metric never changes sign for 2-form fields. This, however, does not mean that hyperbolicity is always preserved. $\bar{g}^{(2)}$ can still attain the problematic value of $0$, and it is now indeed known that this occurs and leads to loss of hyperbolicity. This is demonstrated by decomposing the 2-form field into other fields and examining their time evolution near a configuration that is close to loss of hyperbolicity. Furthermore, it is also known that the eigenvalue structure, hence the whole signature of the effective metric changes discontinuously when $\bar{g}^{(2)}=0$. In loose terms, loss of hyperbolicity occurs when $(\lambda/2) B_{\mu\nu}B^{\mu\nu}$ becomes comparable to $(\lambda/2) B_{\mu\nu} \tilde{B}^{\mu\nu}$, but we note that it can also happen when both terms vanish: recall that $B^2 = B_{\mu\nu}B^{\mu\nu}$ can be zero even when $B_{\mu\nu}$ is nonzero due to the Lorentzian signature. We have not given the details here, but they are similar to what we will outline in Sec.~\ref{higher order self-interaction}, which in turn follows Ref.~\cite{Unluturk2024}.

\section{Self-interacting 1-forms without loss of hyperbolicity}
\label{sec: 1-form loss of hyperbolicity exception}

\subsection{Higher order self interaction}
\label{sec: alternative self-interaction}

We looked at the simplest case of a quartic self-interaction so far, and loss of hyperbolicity occurs for generic self interactions. However, there are also known exceptions. The first case is obtained simply by changing the power of the self interaction term:
\begin{equation}
\label{eq: 1-form sextic potential}
    V = \frac{\lambda m^2}{6}\left(A_\mu A^\mu\right)^3,
\end{equation}
which leads to a very similar calculation, with the change that Eq.~\eqref{eq: 1-form z} is replaced with
\begin{equation}
\label{eq: 1-form z cubic interaction}
z=1 + \lambda (A_\mu A^\mu)^2.
\end{equation}
$z$ plays a similar role as before, e.g. it determines the signature of the effective metric. Namely, in this case the determinant reads as
\begin{equation}
    \bar{g} = - z^{D-1} \left(5z-4\right).
\end{equation}
Hence, for $\lambda>0$, this means that the signature of $\gbar_{\mu\nu}$ never changes since $z$ is always greater than 1. Any odd-powered self interaction term in the Lagrangian leads to a similar result. Moreover, Ref.~\cite{Rubio2024} have performed numerical time evolutions for the sextic potential in Eq.~\eqref{eq: 1-form sextic potential}, and demonstrated that time evolution is always hyperbolic.

\subsection{The massless limit}
\label{sec: massless limit}
Another recent proposal for a well-behaved class of theories considers purely self-interacting fields with vanishing mass, $m=0$~\cite{Banerjee:2025fph}. Our common setup cannot directly check this, since we explicitly included a factor of $m$ in our self-interaction, so, let us consider the theory
\begin{equation}
\label{eq: 1-form Lagrangian 2}
\L = -\frac{1}{4}F_{\mu\nu}F^{\mu\nu} - \frac{1}{4} \tilde{\lambda} (A_\mu A^\mu)^2.
\end{equation}
Similar steps as before lead to a field equation with the principal part
\begin{equation}
\label{eq: massless EOM}
    \bar{\bar{g}}_{\mu\rho} \partial^\mu \partial^\rho A_\nu + \frac{2\tilde{\lambda}}{y} A^\mu A^\rho \partial_\mu F_{\nu\rho}+ \dots
\end{equation}
where
\begin{equation}
    \label{eq: massless effective metric}
    y = \tilde{\lambda} A_\mu A^\mu\ ,\ \bar{\bar{g}}_{\mu\rho} = g_{\mu\rho}+\frac{2\tilde{\lambda}}{y} A_\mu A_\rho
\end{equation}
This is similar to Eq.~\eqref{eq: 1-form eom wavelike form} for this particular coupling, but we used a double-bar for the effective metric to distinguish it from our definition, which we will soon discuss.

Following the same path as before, the principal symbol suggests that some of the modes are again governed by the effective metric $\bar{\bar{g}}_{\mu\rho}$. However, this metric does not change its signature since
\begin{equation}
    \label{eq: massless determinant}
    \det(\bar{\bar{g}}_{\mu\rho}) = \bar{\bar{g}} = g \left(1+ \frac{2\tilde{\lambda} A_\nu A^\nu}{\tilde{\lambda} A_\nu A^\nu} \right) = 3g .
\end{equation}
This is proposed as a possible path to have healthy hyperbolicity for interacting vector fields.

Despite the above reasoning, it is not clear to us that the purely self-interacting theory is well-posed based on this discussion. One crucial part is the second term in Eq.~\eqref{eq: massless EOM}, which diverges even for vanishingly small values of $A_\mu A^\mu$. Ref.~\cite{Banerjee:2025fph} mention that this term vanishes identically in 1+1 dimensions, which is a known fact in the literature~\cite{Coates2022}. However, this is not the case in any other dimension, and it is quite likely that the equations of motion are ill-posed due to this divergent term. This divergence is even more severe than the loss of hyperbolicity we have seen so far, since it occurs even at the smallest field amplitudes.

Our argument can be most easily understood by multiplying Eq.~\eqref{eq: massless EOM} with an overall factor of $y$ so that we have the alternative effective metric
\begin{equation}
\label{eq: massless EOM2}
    \bar{g}_{\mu\rho} \partial^\mu \partial^\rho A_\nu + 2\tilde{\lambda} A^\mu A^\rho \partial_\mu F_{\nu\rho}+ \dots
\end{equation}
where 
\begin{equation}
\label{eq: massless effective metric2}
    \bar{g}_{\mu\rho} = y g_{\mu\rho}+2\tilde{\lambda} A_\mu A_\rho
\end{equation}
Note that this form of the effective metric is a direct analog of what we have used so far, hence the use of the same symbol, $\bar{g}_{\mu\rho}$. Now, we do not have the divergent $1/y$ in the second term of the field equation~\eqref{eq: massless EOM2}, however $\bar{g}_{\mu\rho}$ becomes degenerate when $A_\mu=0$. In other words, either we have the ``well-behaved'' $\bar{\gbar}_{\mu\rho}$ but have a $1/y$ divergence in another term, or remove the $1/y$ divergence but have the ill-behaved effective metric $\bar{g}_{\mu\rho}$. There is a singularity in both pictures.

Overall, we believe that this purely self-interacting vector field theory still suffers from ill-posedness. We will study the vanishing mass case for $p$-form fields in quantitative detail below, and reach similar conclusions.

\section{Self-interacting  higher forms without loss of hyperbolicity}
\label{sec: higher form loss of hyperbolicity exception}

\subsection{Higher order self interaction}
\label{higher order self-interaction}

This is the case where the self interaction potential is generalized to
\begin{equation}
    \label{eq: n self interaction}
    V = \frac{\lambda m^2}{2n} \left(\frac{1}{p!} B^2 \right)^n,
\end{equation}
with $n\geq2$ an integer, which is a further generalization of the sextic potential~\eqref{eq: 1-form sextic potential}. The equation of motion reads
\begin{equation}
\label{eq: p-form field eqn for 2n potential}
    \partial_\alpha M^{\alpha\mu_1\cdots\mu_p} = m^2 z B^{\mu_1\cdots\mu_p},
\end{equation}
where
\begin{equation}
    z = 1 + \lambda\left(\frac{1}{p!}B^2\right)^{n-1},
\end{equation}
and the resulting generalized Lorenz condition is
\begin{equation}
\label{eq: p-form lorenz condition for 2n potential}
    \partial_\alpha (z B^{\alpha\mu_2\cdots\mu_p}) = 0.
\end{equation}

If we proceed as before, the principal symbol is given by
\begin{equation}
    \frac{1}{p!} \mathcal{P}(k)^{\mu_1\cdots\mu_p}_{\phantom{\mu_1\cdots\mu_p}\nu_1\cdots\nu_p} = k^2 \delta^{[\mu_1}_{\nu_1}\cdots\delta^{\mu_p]}_{\nu_p} 
    + \frac{2\lambda\left(n-1\right)}{z\left(p-1\right)!} \left(\frac{1}{p!}B^2\right)^{n-2} k_\alpha k^{[\mu_1|} B^{\alpha|\mu_2\cdots\mu_p]} B_{\nu_1\cdots\nu_p}.
\end{equation}
It is more convenient to condense the long lists of indices into single multi-indices as
\begin{equation}
\label{eq: higher-order interaction principal symbol}
    \mathcal{P}(k)^a_{\phantom{a}b} = k^2 \delta^a_b + \frac{2p\lambda}{z} \left(n-1\right) \left(\frac{1}{p!}B^2\right)^{n-2} C^a B_b,
\end{equation}
where $C^a$ stands for $C^{\mu_1\cdots\mu_p} = k_\alpha k^{[\mu_1|} B^{\alpha|\mu_2\cdots\mu_p]}$. Finally, we obtain the dispersion relation via the determinant as
\begin{equation}
    \det(\mathcal{P}^a_{\phantom{a}b}) = \frac{1}{z} \left(k^2\right)^{d-1} \left(\bar{g}_{\alpha\beta} k^\alpha k^\beta\right),
\end{equation}
with the effective metric
\begin{equation}
\label{eq: higher-order interaction effective metric}
    \bar{g}_{\alpha\beta} = z g_{\alpha\beta} + \frac{2\lambda\left(n-1\right)}{\left(p-1\right)!} \left(\frac{1}{p!}B^2\right)^{n-2} B_\alpha^{\phantom{\alpha}\mu_2\cdots\mu_p} B_{\beta\mu_2\cdots\mu_p}.
\end{equation}
Here, $d=\big(\begin{smallmatrix}
D \\ p
\end{smallmatrix}\big)$ is the number of dimensions of the $p$-form space in $D$ spacetime dimensions. Note that this reduces to Eq.~\eqref{eq: p-form effective metric} for $n=2$. We can now take the determinant of Eq.~\eqref{eq: higher-order interaction effective metric} for $p=2,3$ in $D=4$ dimensions. 

Let us start with a 3-form, where the story is less surprising. The determinant of the metric is
\begin{equation}
    \bar{g}^{(3)} = - \left(1 + \lambda \left(\frac{1}{6}B^2\right)^{n-1}\right) \left(1 + \left(2n-1\right)\lambda \left(\frac{1}{6}B^2\right)^{n-1}\right)^3.
\end{equation}
which, for sextic self-interaction $n=3$, reduces to
\begin{equation}
    \bar{g}^{(3)} = - \left(1 + \lambda \left(\frac{1}{6}B^2\right)^{2}\right) \left(1 + 5\lambda \left(\frac{1}{6}B^2\right)^{2}\right)^3.
\end{equation}
We can see that the determinant never changes sign for $\lambda>0$, just like the 1-form case, which is a very strong indicator that there is no loss of hyperbolicity in this theory, similarly to the case of 1-forms. The same conclusions hold for any odd $n$.

There are surprises for 2-form fields, where the determinant of the effective metric becomes
\begin{equation}
    \label{eq: p=2 interaction n}
    \bar{g}^{(2)} = - \Bigg[z\left(1 + \frac{3}{2}\lambda\left(n-1\right) \left(\frac{1}{p!}B^2\right)^{n-2} B^2\right)
    - \left(\frac{1}{2}\lambda\left(n-1\right) \left(\frac{1}{p!}B^2\right)^{n-2} B \tilde{B} \right)^2 \Bigg]^2,
\end{equation}
where $B\tilde{B} = B_{\mu\nu}\tilde{B}^{\mu\nu}$. In particular, for $n=3$,
\begin{equation}
\label{eq: p=2 interaction 3}
    \bar{g}^{(2)} = - \frac{1}{64} \left[8 + 14\lambda \left(B^2\right)^2 + 3\lambda^2 \left(B^2\right)^4 - 2\lambda^2 \left(B^2\right)^2 \left(B\tilde{B}\right)^2 \right]^2.
\end{equation}

The formulas above preserve the main distinctive property of 2-form fields familiar from the case of $n=2$, see Eq.~\eqref{eq: 2-form det gbar}. That is, Eqs.~\eqref{eq: p=2 interaction n} and~\eqref{eq: p=2 interaction 3} are manifestly non-positive. Recall that this did not mean that the theory was completely healthy for $n=2$. On the contrary, we mentioned that loss of hyperbolicity occurs when $\bar{g}^{(2)}=0$ for $n=2$, and it could occur dynamically starting from healthy field configurations. We will show that the situation is the same for $n=3$. Hence, the self interaction terms in Eq.~\eqref{eq: n self interaction} that can lead to healthy 1-form and 3-form fields, do not have the same effect on 2-form fields. Self-interacting 2-form fields suffer from loss of hyperbolicity for $n=3$ in Eq.~\eqref{eq: n self interaction}.

To demonstrate these points quantitatively, we start with a 1+3 decomposition and look at the field equations in detail. We first introduce, in analogy with electromagnetism, the three vectors
\begin{equation}
    \mathbf{E} = (B_{10}, B_{20}, B_{30}), \quad \mathbf{B} = (B_{23}, B_{31}, B_{12}).
\end{equation}
In this notation we have $B_{\mu\nu}B^{\mu\nu} = 2(\mathbf{B}^2-\mathbf{E}^2)$ and $B_{\mu\nu}\tilde{B}^{\mu\nu} = 4\mathbf{E}\cdot\mathbf{B}$. Similarly introducing $\mathbf{M} = (M_{023}, M_{031}, M_{012})$, Eqs.~\eqref{eq: p-form field eqn for 2n potential} and~\eqref{eq: p-form lorenz condition for 2n potential} can be written as the first order system
\begin{subequations}
\begin{align}
    \dot{\mathbf{B}} &= \mathbf{M} - \nabla\times\mathbf{E}, \\
    \dot{\mathbf{M}} &= \nabla\left(\nabla\cdot\mathbf{B}\right) - m^2 z \mathbf{B}, \\
    \label{eq: E evolution equation}
    \mathbb{A}\dot{\mathbf{E}} &= \nabla\times\left(z\mathbf{B}\right) - 4\lambda\left(\frac{1}{2}B^2\right)\left(\mathbf{B}\cdot{\mathbf{B}}\right)\mathbf{E},
\end{align}
\end{subequations}
subject to the constraint
\begin{equation}
\label{eq: 2n potential constraint equation}
    \nabla\times\mathbf{M} + m^2 z \mathbf{E} = 0,
\end{equation}
where $\mathbb{A} = z\mathbb{I}_3 - 4\lambda\left(\frac{1}{2}B^2\right)\mathbf{E}\mathbf{E}^\mathrm{T}$.

For $\lambda<0$ one can easily see how hyperbolicity can be lost, using the initial configuration
\begin{equation}
    \mathbf{E} = 0, \quad z\mathbf{B} = \nabla\psi, \quad \mathbf{M} = \kappa \nabla\psi,
\end{equation}
where $\kappa$ is a constant and $\psi$ is a scalar function. Note that this satisfies the constraint Eq.~\eqref{eq: 2n potential constraint equation}, and the evolution equations give $\dot{\mathbf{E}} = 0$ and $\dot{\mathbf{B}} = \kappa \nabla\psi$. Therefore, after a small time $\Delta t$, $\mathbf{E}$ remains zero and $\mathbf{B}$ becomes
\begin{equation}
\label{eq: B after small delta t}
    \mathbf{B} = \left(\frac{1}{z} + \kappa \Delta t\right) \nabla\psi.
\end{equation}
With such a configuration the effective metric becomes singular, i.e. the determinant in Eq.~\eqref{eq: p=2 interaction 3} becomes zero for $\lambda (B^2)^2 = \lambda (2\mathbf{B}^2)^2 = -2/3$. But we can choose $|\mathbf{B}| \lesssim (6|\lambda|)^{-1/4}$ by choosing $|\nabla\psi| \lesssim 5 \cdot 6^{-5/4} |\lambda|^{-1/4}$, so that we are close to this singularity. We can then see from Eq.~\eqref{eq: B after small delta t} that we will hit the singularity for large enough $\kappa$.

For $\lambda>0$, it is not as straightforward to find a configuration that loses hyperbolicity, just like the original quartic potential~\cite{Unluturk2024}. One reason for this is that we encounter coordinate singularities, namely, the matrix $\mathbb{A}$ may become non-invertible before we hit the true singularity and therefore time evolution in Eq.~\eqref{eq: E evolution equation} becomes impossible in these coordinates. We expect this to be the case here as well. However, previous results strongly indicate that hyperbolicity is lost for $\lambda>0$ as well if appropriate coordinate changes can be made.\footnote{See Ref.~\cite{Coates2022} and Ref.~\cite{Coates2023} for a more detailed discussion of the coordinate singularity issue in the simpler vector field setting.}

\subsection{The massless limit}
\label{massless p-form}
Recall that there are serious doubts about whether removing the mass term from self-interacting form field theories leads to healthy time evolution. We will further discuss this below, but we will still study this case for the sake of completeness.

The effective metric for the self-interacting $p$-form is
\begin{equation}
    \bar{g}_{\alpha\beta} = zg_{\alpha\beta} + \frac{2\lambda}{\left(p-1\right)!} B_\alpha^{\phantom{\alpha}\mu_2\cdots\mu_p} B_{\beta\mu_2\cdots\mu_p},
\end{equation}
where $z = 1 + (\lambda/p!)B^2$. In order to have a meaningful formula in the massless limit, let us define
\begin{equation}
    \tilde{\lambda} = m^2 \lambda
\end{equation}
as in Eq.~\eqref{eq: 1-form Lagrangian 2}. Furthermore, let us also introduce a new (scaled) effective metric
\begin{equation}
    \tilde{g}_{\mu\nu} = m^2 \bar{g}_{\mu\nu}.
\end{equation}
Note that there is a simple scaling between the two effective metrics, so their hyperbolicity properties are identical. Now, if $m\to0$ while keeping $\tilde{\lambda}$ and $\tilde{g}_{\mu\nu}$ finite,  we obtain
\begin{equation}
    \label{eq: massless p-form effective metric}
    \tilde{g}_{\alpha\beta} = \frac{\tilde{\lambda}}{p!}B^2 g_{\alpha\beta} + \frac{2\tilde{\lambda}}{\left(p-1\right)!} B_\alpha^{\phantom{\alpha}\mu_2\cdots\mu_p} B_{\beta\mu_2\cdots\mu_p},
\end{equation}
which can be used to investigate the time evolution problems. Note that this metric definition is slightly different from that of Eq.~\eqref{eq: massless effective metric}, we have an extra factor of $B^2$. This reflects the issues we discussed about massless $p$-form field theories in Sec.~\ref{sec: massless limit}. $\tilde{g}_{\alpha\beta}$ is simply a scaled version of $\bar{g}_{\mu\rho}$ in Eq.~\eqref{eq: massless effective metric2}. We can remove the $B^2$ from our metric definition to have another effective metric similar to Eq.~\eqref{eq: massless effective metric}, but this only moves the singular nature of the dynamics to another term in the equations of motion. Hence, we will investigate hyperbolicity via Eq.~\eqref{eq: massless p-form effective metric}.

For a 3-form, the determinant of Eq.~\eqref{eq: massless p-form effective metric} is
\begin{equation}
    \tilde{g}^{(3)} = - \frac{1}{48} \left(\tilde{\lambda}B^2\right)^4.
\end{equation}
Thus, this case is similar to the vector case in that we do not have a sign change, and if we removed $B^2$ from the definition of our metric as in Eq.~\eqref{eq: massless effective metric}, the determinant would not depend on the 3-form at all as in Eq.~\eqref{eq: massless determinant}. However, we re-emphasize that there is still a time evolution problem at the risk of verbosity: $\tilde{g}_{\alpha\beta}$ leads to a time evolution problem since it is degenerate when $B^2=0$, and using an alternative effective metric definition without $B^2$ requires dividing the other terms in the equation of motion by $B^2$, again leading to infinities.

Taking the determinant of Eq.~\eqref{eq: massless p-form effective metric} for a 2-form, we get
\begin{equation}
    \tilde{g}^{(2)} = - \tilde{\lambda}^4 \left[3\left(\frac{1}{2}B^2\right)^2 - \left(\frac{1}{2}B\tilde{B}\right)^2 \right]^2,
\end{equation}
The case is less subtle here, and there is no need to consider different definitions of the effective metric. This is because $\tilde{g}^{(2)}$ can be shown to vanish for certain finite values of $B^2$ and $B\tilde{B}$ and lead to loss of hyperbolicity using an argument that is similar to the one in Sec.~\ref{higher order self-interaction}.

In summary, we believe that removing the mass term cannot heal self-interacting $p$-form field theories, the same way it did not heal the 1-form field theories.

\section{Conclusions}
\label{sec: conclusion}
Many aspects of form fields are known to be generalizations of its simplest example, the 1-form field which is directly related to a vector field. This has been known to be the case for loss of hyperbolicity as well, but we also knew that there were significant differences in terms of how loss of hyperbolicity occurs, e.g. between 1-form and 2-form fields in $D=4$ dimensions. Here, we showed that such peculiar differences continue to exist in the case of perfect hyperbolicity as well. The exceptional self-interacting vector field theories with indefinite time evolution have very close analogs in 3-form fields. Yet, there is all the indication that the same form of self interactions still lead to loss of hyperbolicity in 2-form fields.

Moreover, we showed that the proposal to obtain well-posed self-interacting field theories by removing the mass term likely does not work. Such theories seem to have a very simple and healthy time evolution structure for a certain definition of the effective metric, but this is due to moving the singular behavior to other terms and ignoring it, rather than truly removing the singularity. Still, there has not been a detailed analysis of these singular terms that appear in the non-principal parts of the differential equation, hence, a definitive conclusion about the well-posedness of time evolution awaits further studies.

As we mentioned in the beginning, all of our results can be generalized to curved spacetime. Moreover, we expect our later results in $D=4$ to be also generalizable to any spacetime dimension. However, the behavior of 2- and 3-form fields should not be expected to stay analogous in any dimension. The similarities of 1- and 3-form fields is most likely due to the fact that the Hodge dual of a 3-form field is a 1-form field, hence there is an close relationship between the two. Therefore, we would expect that 1-form fields in $D$ dimensions behave similarly to $(D-1)$-form fields. Likewise, in even dimensions, we expect $(D/2)$-form fields to have a distinct hyperbolicity behavior as in the case of $2$-form fields in $D=4$. We naively expect other cases, like $2$-form fields in $D=5$ to behave dissimilarly to 1-form fields, but whether such expectations are valid remains to be seen.

\ack
Metin Gürses is one of the pillars of theoretical physics in Turkey. We express our gratitude for being invited to the celebration of his many decades of research accomplishments.


\section*{References}
\providecommand{\newblock}{}

\end{document}